\documentclass[epj]{svjour}
\usepackage{latexsym,amssymb,amsfonts,color,graphicx,amsmath}
\usepackage{upgreek,bm}
\usepackage[american]{babel}

\usepackage[normal,sc]{caption}
\usepackage{cite}
\begin{document}

\hyphenation{squir-mer squir-mers Pa-ra-me-cia Pa-ra-me-cium}
\newcommand{\mpf}[1]{\begin{center}\fcolorbox{RoyalBlue}{White}{ \begin{minipage}{0.94\textwidth} #1  \end{minipage} }\end{center}}

\newcommand{\eh}{\mathbf{e}}
\newcommand{\deriv}[2]{\frac{\mathrm{d} #1}{\mathrm{d} #2}}
\newcommand{\hvec}[1]{\hat{\mathbf{#1}}}

\newcommand{\tEins}{\textbf{\textsf{1}}}

\newcommand{\Pe}{\mathrm{Pe}}
\newcommand{\Per}{\mathrm{Pe}_r}

\newcommand{\vhh}[1]{\hat{\mathbf{#1}}}
\newcommand{\hsy}[1]{\hat{\boldsymbol{#1}}}
\newcommand{\RCh}{R_{\mathrm{Ch}}}
\newcommand{\RSq}{R_{\mathrm{Sq}}}
\newcommand{\vdd}[1]{\frac{d}{dt}\mathbf{#1}}
\newcommand{\vdh}[1]{\frac{d}{dt}\hat{\mathbf{#1}}}

\newcommand{\hphi}{\hat{\boldsymbol{\varphi}}}
\newcommand{\hrho}{\hat{\boldsymbol{\rho}}}
\newcommand{\hz}{\hat{\mathbf{z}}}
\newcommand{\ephi}{e_\varphi}
\newcommand{\erho}{e_\rho}
\newcommand{\ez}{e_z}
\newcommand{\htheta}{\hat{\boldsymbol{\theta}}}
\newcommand{\vf}{\bar{v}_f}
\newcommand{\HDD}{H_\text{2D}}
\newcommand{\HDDD}{H_\text{3D}}

\newcommand{\refEq}[1]{Eq.~(\ref{Eq:#1})}
\newcommand{\refEqu}[1]{Eqs.~(\ref{Eq:#1})}
\newcommand{\refFig}[1]{Fig.~\ref{Fig:#1}}
\newcommand{\refFigu}[1]{Figs.~\ref{Fig:#1}}
\newcommand{\refFigure}[1]{Figure~\ref{Fig:#1}}
\newcommand{\refFigures}[1]{Figures~\ref{Fig:#1}}
\newcommand{\refSec}[1]{Sec.~\ref{Sec:#1}}
\newcommand{\refChapt}[1]{Chapt.~\ref{Chapt:#1}}
\newcommand{\vecin}[2]{#1 \cdot #2}
\newcommand{\tensA}[1]{\boldsymbol{#1}}

\renewcommand{\ge}{\geqslant}
\renewcommand{\le}{\leqslant}

\newcommand{\nablabf}{\boldsymbol{\nabla}}

\newcommand{\ten}[1]{\vec{\sf #1}}
\newcommand{\NMD}{N_{\text{MD}}}

\newcommand{\hx}{\hat{\vec{x}}}
\newcommand{\hy}{\hat{\vec{y}}}
\newcommand{\const}{\text{const}}

\newcommand{\vout}{\vec{v}_\text{out}}
\newcommand{\vin}{\vec{v}_\text{in}}
\newcommand{\viout}{\vec{v}_\text{i,out}}
\newcommand{\viin}{\vec{v}_\text{i,in}}

\newcommand{\anm}[1]{\textsl{\textcolor{Orange}{#1}}}

\newcommand{\hs}[1]{\textcolor{red}{#1}}
\title{Simulating squirmers with multiparticle collision dynamics}
%
\author{Andreas Z\"{o}ttl\inst{1,2} and Holger Stark\inst{2}
}                     
%
%
\institute{Rudolf Peierls Centre for Theoretical Physics, University of Oxford,
1 Keble Road, Oxford, OX1 3NP, UK  \and Institute for Theoretical Physics,  Technische Universit\"{a}t Berlin, Hardenbergstr.\ 36, 10623 Berlin, Germany \\
\email{andreas.zoettl@physics.ox.ac.uk}
}
\date{Received: date / Revised version: date}
%
\abstract{
Multiparticle collision dynamics is a modern coarse-grained simulation technique
to treat the hydrodynamics of Newtonian fluids 
by solving the Navier-Stokes equations.
Naturally, it also includes
thermal noise. 
Initially it has been applied 
extensively to
spherical colloids or bead-spring polymers immersed in a fluid. 
Here, we review and discuss the use of multiparticle collision dynamics
for studying
the motion of spherical model microswimmers called \textit{squirmers} moving in viscous fluids.%
} 
\maketitle

\section{Introduction}
\label{intro}
Recently, there has been a great interest in understanding the physical principles of the locomotion mechanisms of microswimmers.
Important examples are biological microorganisms such as bacteria, sperm cells or algae, and artificial self-propelled swimmers such as active colloids \cite{Zottl2016,Bechinger2016} or active emulsion droplets \cite{Maass2016}.
Their
locomotion is mainly governed by low Reynolds number hydrodynamics and thermal or biological noise.
While   swimming speed and flow fields of a single noiseless microswimmer in 
a
bulk 
fluid may be calculated analytically 
in some simple cases by solving
Stokes equations \cite{Taylor51,Lighthill52,Blake71,Najafi2004,Avron2005},
analytical methods are less suitable for swimmers in complex environments, 
in the presence of other swimmers, or when Brownian fluctuations become important.
Hence several numerical methods are used nowadays to overcome these limitations.
Very succesful tools are coarse-grained hydrodynamic simulation methods such as 
lattice Boltzmann (LB), dissipative particle dynamics (DPD),
and multiparticle collision dynamics (MPCD) -- 
a special version of which is
stochastic rotation dynamics (SRD).
These methods allow to simulate the 
time evolution
of one or many swimmers, in bulk as well as in confinement.
While MPCD and DPD are particle-based methods and naturally include thermal fluctuations, LB is based on the time
evolution of smooth velocity 
distribution functions
and noise may be included on top.

In particular, MPCD has been used extensively in the last decade to simulate the dynamics and flow fields of 
various nano-and microswimmers like a Taylor sheet \cite{Muench2016},
artificial nano- and microswimmers
\cite{Rueckner07,Tao08,Tao09,Tao09b,Thakur10,Thakur10b,Thakur11,Thakur11b,Yang11,Thakur12,Huang12,deBuyl13,Kapral13,Yang13,Yang2014,Wagner2017},
sperm cells, flagella, cilia, and self-propelled rods \cite{YangB08,Elgeti08,Elgeti09,Elgeti10,YangB10,Reigh12,Reigh13,YangB13,Elgeti13,Yang2014b,Agrawal2018},
 dumbbell swimmers \cite{Schwarzendahl2017},
swimming bacteria \cite{Hu2015,Hu2015b,Zottl2017},
and the complicated motion of the parasite \textit{African Trypanosome} \cite{Babu12b,Heddergott12,Alizadehrad2015} and bacterial swarmer cells \cite{Eisenstecken2016}.
At present a very popular spherical model swimmer is the so-called \textit{squirmer}.
Its
motion has recently
 been modeled 
 with MPCD 
and it has been used to explore generic features of microswimmers
 \cite{Downton09,Goetze10,Zottl2012,Zottl2014,Schaar2015,Blaschke2016,Theers2016,Kuhr2017,Ruehle2018}.
In this paper we will summarize the details how the squirmer is implemented and simulated with MPCD.

We first introduce the basics of the MPCD method in section~\ref{Sec:MPCD} and discuss the squirmer model briefly in section~\ref{Sec:Squirmer}. We then explain in detail how the motion of the squirmer can be coupled to the MPCD fluid using a hybrid molecular dynamics -- multiparticle collision dynamics (MD-MPCD) scheme in section~\ref{Sec:Simulation}.
We then present results for swimming in a narrow slit geometry in section~\ref{Sec:slit},
and finish with a conclusion in section~\ref{Sec:conc}.

\section{Multiparticle collision dynamics}
\label{Sec:MPCD}
Multiparticle collision dynamics (MPCD) is a  coarse-grained solver of the Navier-Stokes equations \cite{Kapral08,Gompper08},
which naturally includes thermal fluctuations because it simulates the fluid
explicitly via pointlike, effective fluid particles 
kept
at  temperature $T_0$.
Transport coefficients such as the shear viscosity $\eta$ 
can be
calculated analytically for a wide range of simulation parameters but 
depend on the particular collision rule between fluid particles \cite{Gompper08}.

The MPCD method was originally  proposed by Malevanets and Kapral and termed \textit{stochastic rotation dynamics} (SRD) \cite{Malevanets99,Malevanets00}. The fluid is modeled by $N_p$ effective  fluid particles, which are pointlike and have
mass $m_0$. 
They
perform alternating \textit{streaming} and \textit{collision} steps
and thereby solve
the Navier-Stokes equations on a coarse-grained level
since the collision step preserves local momentum.
Initially, the particles are randomly placed in the simulation domain at initial positions $\mathbf{r}_i(t=0)$ and at velocities 
 $\mathbf{v}_i(t=0)$ 
that
are drawn from a normal distribution with zero mean and standard deviation $\sigma_0 = \sqrt{k_BT_0/m_0}$.

\begin{figure}[bth]
\begin{center}
\includegraphics[width=.9\columnwidth]{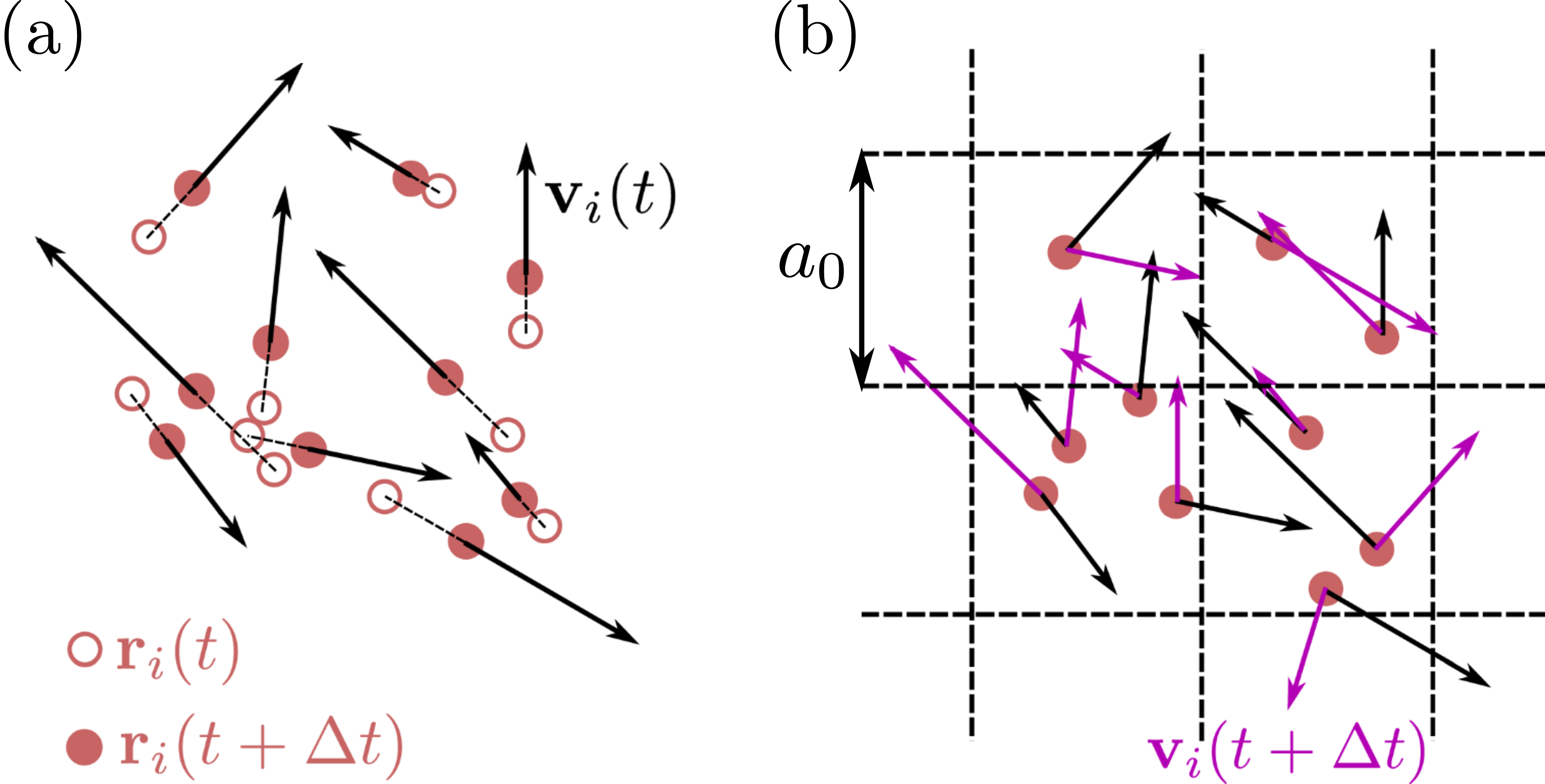}
\end{center}
\caption{Sketch of the MPCD method. (a) Streaming step: Pointlike fluid particles move ballistically for a time $\Delta t$  with 
velocities $\mathbf{v}_i$.
(b) Collision step: Particles are sorted into cubic cells of 
edge
length $a_0$ and 
exchange 
momentum with other particles in the cell
such that the total momentum  is conserved.
 }
\label{Fig:1}
\end{figure}

In the streaming step [see Fig.~\ref{Fig:1}(a)] the fluid particles are moved ballistically for a time $\Delta t$ such that their positions are updated to
\begin{equation}
\mathbf{r}_i(t+\Delta t) = \mathbf{r}_i(t) + \mathbf{v}_i(t)\Delta t.
\label{Eq:str1}
\end{equation}
Noteworthy, here  $\Delta t$ is a simulation parameter, which 
determines
fluid properties such as the viscosity.
This is in contrast to molecular dynamics (MD) simulations, where $\Delta t$ only determines the accuracy 
of a
simulation
run. Typically, $\Delta t = 10^{-2}-10^{0}$ in MPCD units.
After the streaming step all the fluid particles are sorted into cubic cells 
with edge
length $a_0$.
To perform
the collision step, all particles in a cell interact with each other and their velocities are updated  to [see also \refFig{1}(b)]
\begin{equation}
\mathbf{v}_i(t+\Delta t) = \mathbf{u}_\xi(t) + \boldsymbol{\Xi}(\mathbf{r}_j(t),\mathbf{v}_j(t)).
\label{Eq:coll000}
\end{equation}
Here $\xi=1,\dots,N_c$ ist the cell index, $N_c$ is the total number of cells, and
 $\mathbf{u}_\xi(t) $ is the average 
 particle
 velocity in cell $\xi$.
The collision operator
$ \boldsymbol{\Xi}(\mathbf{r}_j(t),\mathbf{v}_j(t)) $  in general depends
on  positions and velocities of the fluid particles in cell $\xi$.
It
has to be chosen such that momentum
is conserved locally in each cell, in order to reproduce
a flow field, which solves
the Navier-Stokes equations.
Depending on its concrete realization it either conserves energy or keeps temperature and thus the mean kinetic
energy constant.

In the original SRD method the fluid particles are part of  a 
microcanonical
ensemble and both momentum and energy is conserved locally. 
The
collision operator reads
\begin{equation}
\boldsymbol{\Xi} = \mathcal{R}_\alpha(\vec{v}_i-\vec{u}_{\xi}),
\end{equation}
where $\mathcal{R}_\alpha$ is a rotation operator, which rotates the relative velocities $\vec{v}_i-\vec{u}_{\xi}$
about a randomly chosen axis by a fixed angle $\alpha$. 
In order to fulfill Galilean invariance but also to reduce memory effects and correlations in the collision step,
the cell grid is shifted randomly
at each
time step \cite{Ihle03} by a random vector $\mathbf{s}=\{s_1,s_2,s_3\}$, where the 
components
$s_i$ are drawn from a uniform distribution $s_i \in [-a_0/2,a_0/2]$.

In their original paper Malevanets and Kapral showed  \cite{Malevanets99} that by averaging out 
the particle dynamics on short length and time scales using
a Chap\-man-Enskog expansion of the corresponding Boltzmann equation,  the Navier-Stokes equations can be derived.
Furthermore, the
$\mathcal{H}$-theo\-rem guarantees fluid relaxation towards equilibrium.
The authors
also demonstrated that the number of fluid particles per cell  is 
Poisson 
distributed
and 
the
fluid particle velocity distribution
converges
fast towards 
a
Maxwell-Boltzmann 
distribution.
In addition, transport coefficients like 
the fluid viscosity can be calculated from the relevant model parameters $m_0$, $a_0$, $T_0$, $\gamma$, $\alpha$ and $\Delta t$ \cite{Malevanets99,Malevanets00,Ihle01,Ihle03,Tuezel03,Pooley05,Tuezel06},
where $\gamma=\sum_{\xi=1}^{N_c} n_\xi / N_c$ is the average number of fluid particles $n_\xi$ per cell.

Noguchi \emph{et al.} developed alternative collision operators for MPCD fluids.
They
use a Langevin or Anderson thermostat 
to 
keep
temperature
constant
but also 
to
conserve momentum
locally
\cite{Noguchi07,Noguchi08}.
Thus,
a canonical ensemble of fluid particles is modeled,
where energy
is 
conserved on average.
An advantage of 
these alternative
collision rules is that local angular momentum conservation,
which is missing in the original SRD formulation, can be included in a rather simple way.
Although angular momentum conservation does not directly alter the Navier Stokes equations,
it ensures that the stress tensor is symmetric,
as it should be for the
Newtonian 
fluid with its isotropy. Indeed, it
has 
been shown that neglecting angular momentum conservation in the fluid may lead to 
incorrect physical effects \cite{Goetze07}.
One popular approach
for the collision operator
is to use the Anderson thermostat 
and, in addition, conserve
angular momentum. 
This type of collision rule is
abbreviated by \textit{MPC-AT+a} \cite{Noguchi07}.
It
has
widely
been used 
to study the dynamics of squirmers in MPCD fluids 
\cite{Goetze10,Zottl2012,Zottl2014,Schaar2015,Blaschke2016,Kuhr2017,Ruehle2018}.

\section{Squirmer model}
\label{Sec:Squirmer}

\begin{figure}[htb]
\includegraphics[width=\columnwidth]{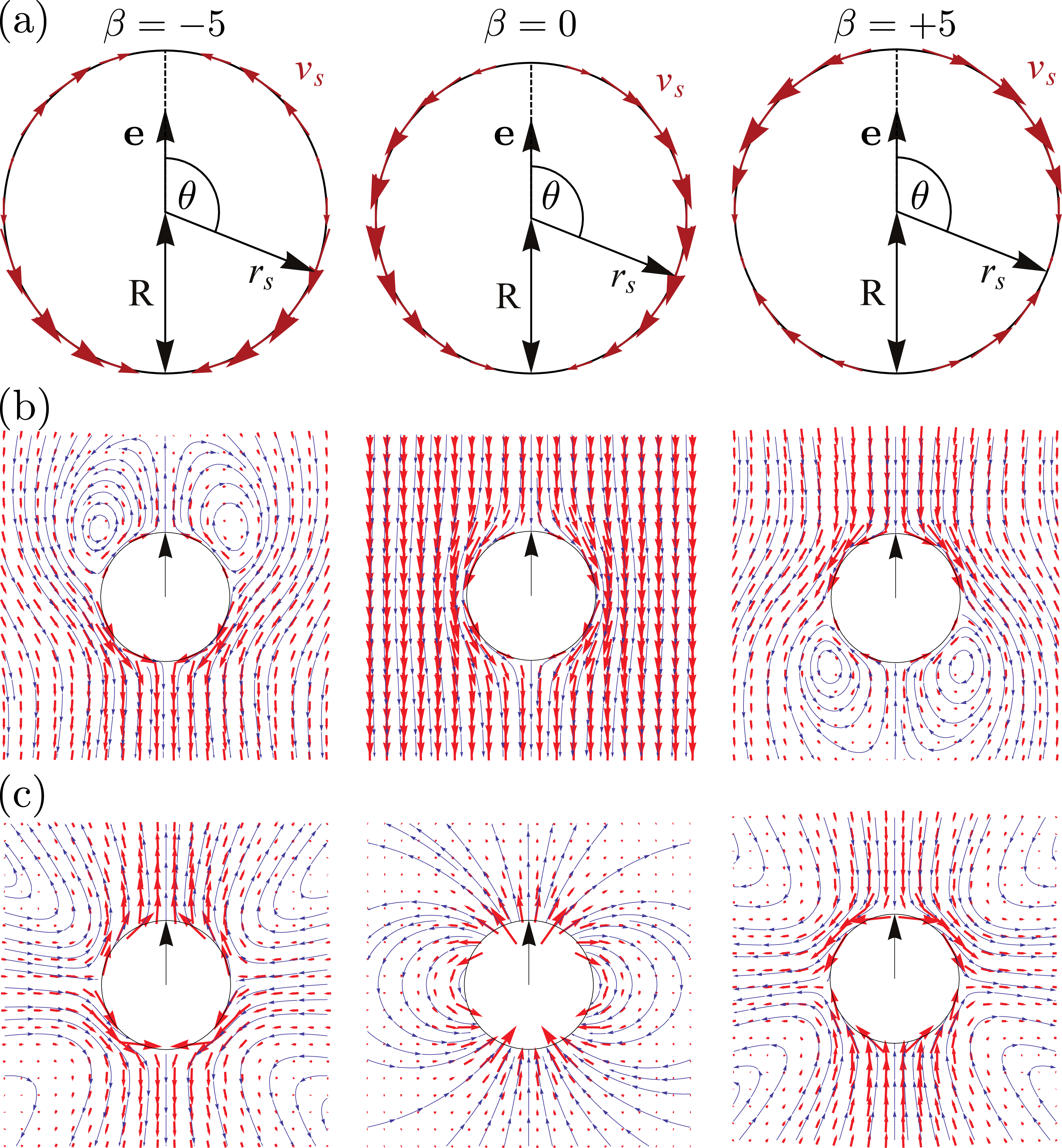}
\caption{
The squirmer model.
(a) Axisymmetric surface velocity profile $\mathbf{v}_s(\theta)$ for a pusher ($\beta=-5$),
a neutral squirmer ($\beta=0$) and a puller ($\beta=+5$). 
(b) Corresponding velocity fields (red) and stream lines (blue) in the frame moving with the swimmer.
(c)  Velocity fields (red) and stream lines (blue) in the lab frame.
}
\label{Fig:Squirmer1}
 \end{figure}

The squirmer is a simple spherical model swimmer.
In its simplest form, a single parameter allows to tune the flow field initiated by the squirmer in the surrounding fluid. Thereby it mimics
typical types of microswimmers ranging from \textit{pushers} via neutral swimmer to \textit{pullers}. The squirmer
has been introduced originally  as a model  for  
ciliated microorganisms such as
the spherical
\textit{Volvox} or 
the elongated
\textit{Opalina} \cite{Lighthill52,Blake71}.
Their cell 
bodies are
covered entirely  by  synchronously beating cilia, which propel them forward
by propagating so-called metachronal waves along the cell surface
 \cite{Brennen77}.
After averaging over a full beating cycle, characterized by a \textit{power stroke} and \textit{recovery stroke}, the resulting  net flow close to the surface is nonzero.
To account for these flows,
the squirmer model was developed by Lighthill \cite{Lighthill52} and extended  by Blake \cite{Blake71}.
Cilia-induced flows close to the surface of the swimmer is here modeled by an effective slip velocity  $\mathbf{v}_s $ directly at 
the surface of the sphere. Considering only axisymmetric flows, the time-dependent surface 
velocity fields in spherical coordinates
read
\begin{equation}
 \mathbf{v}_s(\theta,t) = v_\theta(\theta,t)\mathbf{e}_{\theta} + v_r(\theta,t)\mathbf{e}_r.
\end{equation}
Without any thermal noise, a
squirmer in bulk simply swims along its  symmetry axis $\mathbf{e}$, and $\theta$ is the angle between
this axis and the point $\vec{r}_s$ on the surface [see Fig.~\ref{Fig:Squirmer1}(a)].
The surface velocities can be expanded in Legendre polynomials $P_n$,
\begin{equation}
\begin{split}
 v_\theta(\theta,t) =& \sum_n A_n(t)P_n(\cos\theta) \\
 v_r(\theta,t) =& \sum_n B_n(t)V_n(\cos\theta),
\end{split}
\label{Eq:SqBC}
\end{equation}
where
\begin{equation}
V_n(\cos\theta) = \frac{2\sin\theta}{n(n+1)}\deriv{P_n(\cos\theta)}{\cos\theta}.
\end{equation}
The expansion coefficients $A_n$ and $B_n$  are  sometimes called \textit{surface velocity modes}.
Solving the governing equations of low-Reynolds-number hydrodynamics, the Stokes equations \cite{Kim,Happel},
together with the  boundary conditions [\refEq{SqBC}]
gives analytic expressions for the flow fields created by a squirmer in the surrounding fluid. 

In many cases the surface velocity is  assumed to be only tangential to the squirmer surface and 
time-indepen\-dent. 
This describes in a good approximation the hydrodynamic flows close to spherical artificial microswimmers.
Examples are active colloids \cite{Zottl2016,Bechinger2016} and droplets \cite{Schmitt13,Schmitt2016,Maass2016}
which move themselves forward realizing  static tangential stresses close to their surface,
for example induced by self-phoretic mechanisms or self-induced Marangoni flows, respectively.
Other examples are biological microswimmers such as
\textit{Paramecia}  \cite{Ishikawa06b} or \textit{Volvox} \cite{Pedley2016}
which propel themselves by synchronously beating thousands of tiny cilia attached to their surface.
Averaged over a beating cycle, they induce
 strong  tangential flows to move forward.
Setting $v_r(\theta,t)=0$, the  velocity field $\mathbf{v}(\mathbf{r})$
around a squirmer of radius $R$ with orientation $\eh$ 
and
located at position $\mathbf{r}_0$ can be calculated to \cite{Ishikawa06}
\begin{equation}
\begin{split}
 & \mathbf{v}(\mathbf{r}) =  -\frac 1 3 \frac{R^3}{r^3}B_1 \eh + \frac{R^3}{r^3}B_1(\vecin{\eh}{\hvec{r}})\hvec{r} \\
 & + \sum_{n=2}^{\infty} \left( \frac{R^{n+2}}{r^{n+2}}
 - \frac{R^{n}}{r^{n}}\right)B_n P_n  (\vecin{\eh}{\hvec{r}})\hvec{r}        \\
 & + \sum_{n=2}^{\infty} \left( \frac n 2 \frac{R^{n+2}}{r^{n+2}} - \left(\frac n 2 -1  \right)\frac{R^{n}}{r^{n}} 
   \right)B_nW_n(\vecin{\eh}{\hvec{r}}) \left[(\vecin{\eh}{\hvec{r}})\hvec{r}-\eh  \right].
\end{split}
\label{Eq:EQS1}
\end{equation}
Here
$r=|\mathbf{r}-\mathbf{r}_0|$ is the distance from the 
squirmer center,
$\hat{\mathbf{r}} = (\mathbf{r}-\mathbf{r}_0)/r$ is the radial unit vector,
and
\begin{equation}
W_n(\vecin{\eh}{\hvec{r}}) = W_n(\cos\theta) = \frac{2}{n(n+1)}\deriv{P_n(\cos\theta)}{\cos\theta} 
\end{equation}
with $ \vecin{\eh}{\hvec{r}} = \cos\theta$.

The swimming speed $v_0$ of a noiseless squirmer in bulk  solely depends on the first mode $B_1$ \cite{Lighthill52},
\begin{equation}
v_0 = \frac 2 3 B_1,
\label{Eq:v0}
\end{equation}
and its velocity is constant, $\vec{V}_0=v_0\eh$.
It can be shown that
 the velocity of a sphere with 
tangential surface velocity  $\vec{v}_s$
is equivalent to the average velocity on the surface \cite{Anderson1989,Stone96},
\begin{equation}
\vec{V}_0 = -\frac{1}{4\pi R^2}\int_S \vec{v}_s \mathrm{d}S.
\end{equation}
Indeed, it
can be shown directly that
only
the first mode gives a nonzero contribution to the integral resulting in
\refEq{v0}.

It is interesting to have a closer look 
at
the flow field presented in \refEq{EQS1}, in particular 
at
the behavior far away from the 
squirmer.
The flow field created by the first mode $B_1$ 
is that of a \textit{source dipole}.
It 
decays as $\sim r^{-3}$ 
and
is often 
the dominant contribution for
spherical microswimmers \cite{Zottl2016}.
The flow field created by the second mode $B_2$
is  a \textit{force dipole} field.
It
decays as $\sim r^{-2}$ and is the dominant field far away from the 
squirmer,
because all 
other
modes $B_i$
with
$i\ge 3$ decay faster, see \refEq{EQS1}.
For $B_2<0$ the far-field is the same as for a so-called pusher (or \textit{contractile} swimmer) and for $B_2>0$ a so-called puller
(or \textit{extensile} swimmer)
is realized.
Typical pushers 
have 
the
propelling apparatus at the back of their cell 
bodies, as observable
in
many bacteria or sperm cells.
In contrast, pullers like \textit{Chlamydomonas} have their flagella in 
front
of the
 cell body.
By tuning $B_2$, the squirmer can therefore be used as a simple model to capture different types of swimmers:
 pushers ($B_2<0$), pullers ($B_2>0$), and source-dipole swimmers ($B_2=0$). 

In order to capture the 
flow fields of microswimmers
in leading order
correctly,
it is sufficient to take the  first two modes $B_1$ and $B_2$ into account.
Experiments with active  emulsion droplets have indeed demonstrated that the flow field can be approxiamted by these two modes only  \cite{Thutupalli11}.
The flow field of a swimming \textit{Paramecium} has been measured as well.
It
can also be approximated well by considering only the 
leading
modes
with
$B_1$ 
being the dominant mode
\cite{Ishikawa06b}.

Therefore, 
in numerical simulations
the squirmer is very often used
in 
its simplest
form, only taking the first two modes into account.
The resulting
surface velocity 
field
reads
\begin{equation}
 \mathbf{v}_s(\mathbf{r}_s,\eh)=
  B_1 \left(1 + \beta(\vecin{\eh}{\hat{\mathbf{r}}_s})    \right)  \left[ (\vecin{\eh}{\hat{\mathbf{r}}_s})\hat{\mathbf{\mathbf{r}}}_s -\eh  \right] \, .
 \label{Eq:vs1}
\end{equation}
Here,
$\hat{\mathbf{r}}_s = \mathbf{r}_s/R$ is the radial unit vector pointing from the squirmer center to its surface
and the \textit{squirmer parameter} $\beta$,
\begin{equation}
\beta= \frac{B_2}{B_1},
\label{Eq:beta}
\end{equation}
quantifies
the leading-order flow field: pushers ($\beta<0$), pullers ($\beta>0$), and source-dipole swimmers ($\beta=0$).
Figure\ \ref{Fig:Squirmer1}(a)
shows a sketch of  squirmers and their surface velocity 
fields
for $\beta=-5$, $\beta=0$, and $\beta=+5$. Its dimensionless velocity field  $\mathbf{v}(\mathbf{r},\eh)/B_1$
generated in the surrounding fluid
can then be written as
\begin{equation}
\begin{split}
\frac{\mathbf{v}(\mathbf{r})}{B_1} =&  
\frac{R^3}{r^3}   \left(  (\vecin{\eh}{\hat{\mathbf{r}}})\hat{\mathbf{r}} - \frac 1 3 \eh   \right) \\
&+\beta  \left( \frac{R^4}{r^4}-\frac{R^2}{r^2} \right)\frac 1 2 \left( -1+3(\vecin{\eh}{\hat{\mathbf{r}}})^2 \right)\hat{\mathbf{r}} \\
&+\beta\frac{R^4}{r^4} (\vecin{\eh}{\hat{\mathbf{r}}})\left[ (\vecin{\eh}{\hat{\mathbf{r}}})\hat{\mathbf{r}} - \eh    
     \right],
\end{split}
\label{Eq:VFieldSq1}
\end{equation}
In
leading order a force dipole field
occurs,
\begin{equation}
\frac{\mathbf{v}_D(\mathbf{r})}{B_1} 
 = -\frac{\beta}{2} \frac{R^2}{r^2} \left[ -1+3(\eh\cdot\vhh{r})^2  \right]\vhh{r},
 \label{Eq:DipoleSq} 
\end{equation}
where the strength of the field is set by the squirmer parameter $\beta$ multiplied with the surface 
area
of the sphere $\sim R^2$.
Similarly, the  mode $B_1$ is related to the source-dipole,
\begin{equation}
\frac{\mathbf{v}_{SD}(\mathbf{r})}{B_1} =
\frac{R^3}{r^3}   \left(  (\vecin{\eh}{\hat{\mathbf{r}}})\hat{\mathbf{r}} - \frac 1 3 \eh   \right) \, ,
 \label{Eq:SDipoleSq} 
\end{equation}
which scales with the volume of the squirmer
$\sim R^3$.
\refFigures{Squirmer1}(b)
and
(c)  show flow fields (red) and stream lines (blue) around pushers ($\beta<0$), pullers ($\beta>0$), and neutral squirmers 
($\beta=0$)
both
in the swimmer (b) and the lab frame (c).

Building on the pioneering work of Lighthill \cite{Lighthill52} and Blake \cite{Blake71}, the squirmer was rediscovered by Pedley 
and co-workers.
They 
used
the squirmer  to model the nutrient uptake of spherical microswimmers \cite{Magar03,Magar05},
the flow around swimming \textit{Volvox} \cite{Pedley2016},
the pairwise hydrodynamic interaction between two squirmers and between two \textit{Paramecia} \cite{Ishikawa06,Ishikawa06b,Giacche10},
collective motion of squirmers in bulk \cite{Ishikawa07b,Ishikawa07c,Ishikawa08b,Ishikawa10,Evans11}, in flow \cite{Ishikawa07a,Ishikawa12,Pagonabarraga13}, and in a monolayer \cite{Ishikawa08a}. Within the last ten years the squirmer 
was also used 
extensively
by other groups  to study the persistent random walk of Brownian squirmers \cite{Downton09},
optimal swimming efficiency and feeding \cite{Michelin10,Michelin11},
finite Reynolds number locomotion \cite{Wang12a,Khair2014,Chisholm2016,Li2016},
unsteady squirming \cite{Wang12b}, the motion in a stratified or unsteady fluid \cite{Doostmohammadi12,Ishimoto13},
propulsion in complex fluids \cite{Leshansky09,Zhu11,Zhu12,Datt2015,DeCorato2015,Nganguia2017}, fluid mixing \cite{Lin11,Pushkin12},
collective motion  \cite{Aguillon12,Molina13,Alarcon13,Matas-Navarro2014,Zottl2014,Kyoya2015,Oyama2016,Delfau2016,Yoshinaga2017,Alarcon2017}, the motion in bounded flows \cite{Pagonabarraga13},
motion under gravity \cite{Kuhr2017,Ruehle2018},
motion in obstacle lattices \cite{Chamolly2017},
alignment in nematic fluids \cite{Lintuvuori2017},
motion near surfaces or in channels \cite{Llopis10,Spagnolie12,Wang13,Zhu13,Ishimoto13b,Li2014,Schaar2015,Papavassiliou2015,Lintuvuori2016,Simmchen2016,Shen2017} and squirmers in optical traps \cite{Navarro10}.

\section{Details of MPCD simulations of squirmers}
\label{Sec:Simulation}
In this section we discuss in detail how the dynamics of squirmers
moving
in 
the
MPCD 
fluid 
is implemented.
We also 
explain how
the dynamics of multiple 
squirmers is treated,
how to include solid boundaries, and how pressure-driven channel flow is 
realized
in the simulation.

In the streaming step both squirmers and fluid particles are moved
forward. In addition, squirmers collide with each other and squirmers and fluid particles hit
bounding walls.
In the collision step
\textit{virtual particles} contribute when fluid particles exchange momentum. They are
inside squirmers and walls that overlap with collision cells.
In both  streaming and collision step momentum and angular momentum are transferred between fluid, squirmers, and walls.
In the following we describe
in detail the
procedure how to perform the actual simulations.

Typical simulation 
geometries in 3D are bulk systems 
implemented by
periodic boundary conditions in $x$, $y$, and $z$ direction,
or
a
 fluid confined between two infinitely extended parallel walls separated by a distance $H$. 
For the latter case the number of collision cells in the direction perpendicular to the walls is 
$H/a_0+1$. This ensures that despite the random grid shifts
all fluid particles and a sufficiently thick layer of virtual particles are covered by the grid all the time.
Initially, the effective fluid particles are randomly dis\-tri\-buted in the simulation domain but are not allowed to overlap with the squirmers.
The velocities are normal distributed with zero mean and standard deviation $\sigma_0 = \sqrt{k_BT_0/m_0}$.
The mass density of the fluid is given by $\rho_f = \gamma m_0/a_0^3$ and the number density by
$n_f = \gamma/a_0^3$ with $\gamma$ the average number of fluid particles per cell.

One possibility is to choose cell length $a_0$, mass $m_0$, and energy $k_BT_0$ as the  basic units \cite{Padding06}.
Then
the unit of time 
becomes
 $\tau_0=a_0\sqrt{m_0/k_BT_0}$ and the unit of viscosity is $\eta_0 = \sqrt{m_0 k_BT_0}/a_0^2$.
The remaining parameters are
$\Delta t$, 
$\gamma$, and,
in addition,
the rotation angle $\alpha$, when the SRD collision operator is used.
These parameters  determine  transport coefficients of the fluid such as the viscosity 
\cite{Gompper08}.

In our simulations 
we consider the dynamics of one or $N_S$
identical squirmers
of radius $R$, 
with
surface velocity modes 
$B_1$ and $B_2$, 
and with
the same mass density as the fluid, $\rho_S=\rho_f$. Note, when  $B_1=B_2=0$ passive colloids 
can be simulated as well. Squirmers are initially placed such that they do not overlap with other squirmers, walls, or fluid particles.
The 
concrete
initial conditions 
for
their positions and orientations depend on the specific problem of interest.
Usually, initial values for velocities and angular velocities are set to  $\mathbf{V}_i=0$ and  $\boldsymbol{\Omega}_i=0$.
The mass $M_S$ of a squirmer is given by
\begin{equation}
M_S = \rho_SV_S= \frac{4\pi\gamma  m_0  R^3}{3 a_0^3}
\end{equation}
with volume $V_S=4\pi R^3/3$.
The moment of inertia tensor of a  squirmer is isotropic, $\tens{I}_S = I_S\tEins$, with
\begin{equation}
I_S = \frac{2M_S R^2}{5} = \frac{8\pi\gamma R^5}{15} \frac{m_0}{a_0^3}.
\end{equation}

\subsection{Streaming step}
In the streaming step  squirmers and effective fluid particles move ballistically and 
they
interact with each other and with confining walls. There are different ways how to integrate the streaming of the system.
One way is to use 
interaction
potentials, such as a cut-off Lennard-Jones
or WCA
potential \cite{Weeks71},  
between squirmers 
to ensure they do not overlap.
Then, the streaming step is divided into $\NMD$ time steps, which defines the MD time interval $\delta t = \Delta t / \NMD$.
It
has to be chosen sufficiently small in order to correctly integrate 
motion in
possibly 
very steep potentials.
Another
way is to perform event-driven MD simulations 
during the time intervall
$\delta t = \Delta t$.

\paragraph{Motion of the squirmers}

Squirmers are moved ballistically 
during
time $\delta t$ in order to update their positions $\mathbf{R}_i$ and orientations $\eh_i$.
In some situations external forces $\mathbf{F}_i(t)$ 
and/or
torques $\mathbf{T}_i(t)$ 
act on
the squirmers.
They can also include the interaction forces between squirmers.
Then, one
possibility
to move squirmers in the streaming step
is to use a simple \textit{Velocity-Verlet} integration  \cite{Verlet67,Swope82}.
It updates positions
and orientations 
according
to
 \begin{equation}
 \begin{split}
   \mathbf{R}_i(t+\delta t) &= \mathbf{R}_i(t)+ \left[ \mathbf{V}_i(t) + \frac{1}{2M_S} \mathbf{F}_i(t)|\delta t|  \right] \delta t, \\
   \eh_i(t+\delta t) &= \eh_i(t)+\left[ \left( \boldsymbol{\Omega}_i(t) + \frac{1}{2I_S} \mathbf{T}_i(t)|\delta t| \right)\times\eh_i(t)\right] \delta t,
 \end{split}
\label{Eq:dr2}
\end{equation}
and translational and angular velocities 
according
to
 \begin{equation}
 \begin{split}
   \mathbf{V}_i(t+\delta t) &= \mathbf{V}_i(t)+ \frac{1}{2M_S}\left[ \mathbf{F}_i(t)+ \mathbf{F}_i(t+\delta t)\right] \delta t, \\
   \boldsymbol{\Omega}_i(t+\delta t) &= \boldsymbol{\Omega}_i(t)+ \frac{1}{2I_S}\left[ \mathbf{T}_i(t)+ \mathbf{T}_i(t+\delta t)\right] \delta t.
 \end{split}
\label{Eq:dr3}
\end{equation}
When hard potentials between squirmers are used, an
 alternative is to perform
event-driven simulations. 
Squirmers
may overlap when they move for a time $\delta t$.
Then, all
collision times $t_{ij}$  between 
squirmer $i$ and 
another squirmer $j$ or a wall $j$
have to be calculated.
Instead of moving 
during time
$\delta t$,
 the squirmers
can only move for a time $t_m <  \delta t$ with   $t_m = \min (t_{ij})$.
If squirmer $i$ hits a wall and the wall is assumed to be smooth 
so that it does not act with a frictional torque on the squirmer,
the squirmer
velocity is 
simply updated to
\begin{equation}
 \mathbf{V}_i'= \mathbf{V}_i - 2  (\vecin{\mathbf{V}_i}{\mathbf{n}})\mathbf{n},
\end{equation}
with $\mathbf{n}$ the wall normal.
If two squirmers $i$ and $j$
are the
collision partners,
which have smooth surfaces so that they do not exchange angular momentum,
their velocities are updated to \cite{Allen87}
 \begin{equation}
 \begin{split}
  \mathbf{V}_i' &= \mathbf{V}_i + \delta \mathbf{V}_{ij}, \\
  \mathbf{V}_j' &= \mathbf{V}_j - \delta \mathbf{V}_{ij},
 \end{split}
\label{Eq:dr4}
\end{equation}
with
\begin{equation}
 \delta \mathbf{V}_{ij} =
 -\frac{1}{4R^2}  \left( \vecin{\mathbf{R}_{ij}}{\mathbf{V}_{ij}} \right)  \mathbf{R}_{ij}
\end{equation}
and
\begin{equation}
\begin{split}
 \mathbf{R}_{ij} &= \mathbf{R}_{i} - \mathbf{R}_{j}, \\
 \mathbf{V}_{ij} &= \mathbf{V}_{i} - \mathbf{V}_{j}. \\
\end{split}
\end{equation}
After the collision is performed,
all squirmers are moved 
during time interval $\delta t - t_m$ and again the collision partners and times are determined and their velocities updated.
This procedure is repeated  until the total time $\delta t$ is consumed.

\paragraph{Motion of the fluid particles}
In the absence of external forces acting on the fluid and fluid-squirmer interactions,
the streaming of the fluid particles is simply given by \refEq{str1}.
However, to implement a Poiseuille flow in a channel geometry or between two infinitely extended parallel plates,
one needs to apply a constant force $\mathbf{f}_i$ to them representing the constant pressure gradient \cite{Allahyarov02}.
Then, again the Velocity-Verlet algorithm can be used to update positions and velocities,
\begin{equation}
\begin{split}
\mathbf{r}_i(t+\delta t) &= \mathbf{r}(t) + \mathbf{v}_i(t)\delta t + \frac{1}{2m_0} \mathbf{f}_i |\delta t| \delta t, \\
\mathbf{v}_i(t+\delta t) &= \mathbf{v}_i(t) + \frac{1}{2m_0}\left[  \mathbf{f}_i(t) + \mathbf{f}_i(t+\delta t) \right]\delta t.
\end{split}
\label{Eq:StrF}
\end{equation}

Fluid particles also collide with squirmers or bounding walls during the streaming step.
In contrast to the 
event-driven implementation of
squirmer dynamics, 
one 
may 
 not want to  calculate exact collision times between fluid particles and squirmers or walls
since this slows down the simulation.
One way to overcome this is to first stream 
all
the squirmers
during time $\delta t$ neglecting that some
fluid particles may enter 
them and only then move the fluid particles.
Padding et al.~\cite{Padding05} have proposed an efficient method to do so for simulating colloidal
suspensions, which has been
used for squirmer-fluid interactions as well \cite{Downton09,Zottl2012,Zottl2014,Schaar2015}.
We start with a fluid particle hitting a wall. Once this happens,
it is moved half a time step back and its velocity is updated
as follows.
In order to fulfill the no-slip boundary condition at the wall, the bounce back rule is applied, 
where the velocity is simply reversed,
\begin{equation}
\mathbf{v}_i' = - \mathbf{v}_i \, ,
\label{Eq:fluidIA1}
\end{equation}
meaning that surfaces are rough for the fluid.
Then, the fluid
particle is moved forward for half a time step with the new velocity.
A similar procedure is applied if fluid particle $i$ hits squirmer $j$,
but one has to account for the fact that the surface of the squirmer 
carries
a 
slip
velocity
field
as well  \cite{Downton09},  
\begin{equation}
  \mathbf{v}_i' = -\mathbf{v}_i + 2\left[ \mathbf{v}_s(\eh_j,\mathbf{r}_{ij}^\ast)
  + \boldsymbol{\Omega}_j \times (\mathbf{r}_{ij}^\ast-\mathbf{R}_j) + \mathbf{V}_j \right] \, .
\label{Eq:fluidIA2}
\end{equation}
Here,
$\mathbf{r}_{ij}^\ast$ is the collision point of the particle on the surface of the squirmer
and $ \mathbf{v}_s(\eh_j,\mathbf{r}_{ij}^\ast)$ is the 
slip velocity
[\refEq{vs1}] of  squirmer $j$ located at  $\mathbf{r}_{ij}^\ast$.
When
passive colloids are simulated,
$\mathbf{v}_s$ is set to zero.
Note that 
during  time interval  $\delta t$
a fluid particle is allowed to interact several times
with different squirmers or 
bounding walls.
These
multiple reflections help to reduce possible depletion interactions between squirmers.

When a fluid particle interacts with  a squirmer or a wall, its  momentum $\mathbf{p}_i=m_0\mathbf{v}_i$ is modified.
While a fixed wall absorbs this momentum, moving objects such as colloids or squirmers 
update their velocity and angular velocity such that the total momentum and angular momentum is conserved during the collision.
The change of momentum for fluid particle $i$ hitting  squirmer $j$ is simply 
\begin{equation}
  \Delta \mathbf {p}_{ij}=m_0[ \mathbf{v}_i' - \mathbf{v}_i ] \, ,  
\end{equation}
which is then transferred to the squirmer.
In general,
if $N_j^\ast$ fluid particles hit
squirmer $j$ 
during
time interval $\delta t$,
its 
velocity
$\mathbf{V}_j$  and angular 
velocity
$\boldsymbol{\Omega}_j$  are updated  to
 \begin{equation}
 \begin{split}
   \mathbf{V}_j' &= \mathbf{V}_j +  \frac{1}{M_S}  \Delta \mathbf{P}_j^s, \\
   \boldsymbol{\Omega}_j' &= \boldsymbol{\Omega}_j +\frac{1}{I_S}   \Delta \mathbf{L}_j^s,
 \end{split}
\end{equation}
where
 \begin{equation}
 \begin{split}
  \Delta \mathbf{P}_j^s &= -\sum_{i=1}^{N_j^\ast}\Delta \mathbf{p}_{ij}, \\
  \Delta \mathbf{L}_j^s &= - \sum_{i=1}^{N_j^\ast} \left[ ( \mathbf{r}_{ij}^\ast - \mathbf{R}_j) \times \Delta\mathbf{p}_{ij} \right].
 \end{split}
\label{Eq:dPStr}
\end{equation}
are
the total  linear and angular momentum 
transferred to  squirmer $j$ in
the streaming step.

\subsection{Collision step}
\begin{figure}[bth]
\begin{center}
\includegraphics[width=\columnwidth]{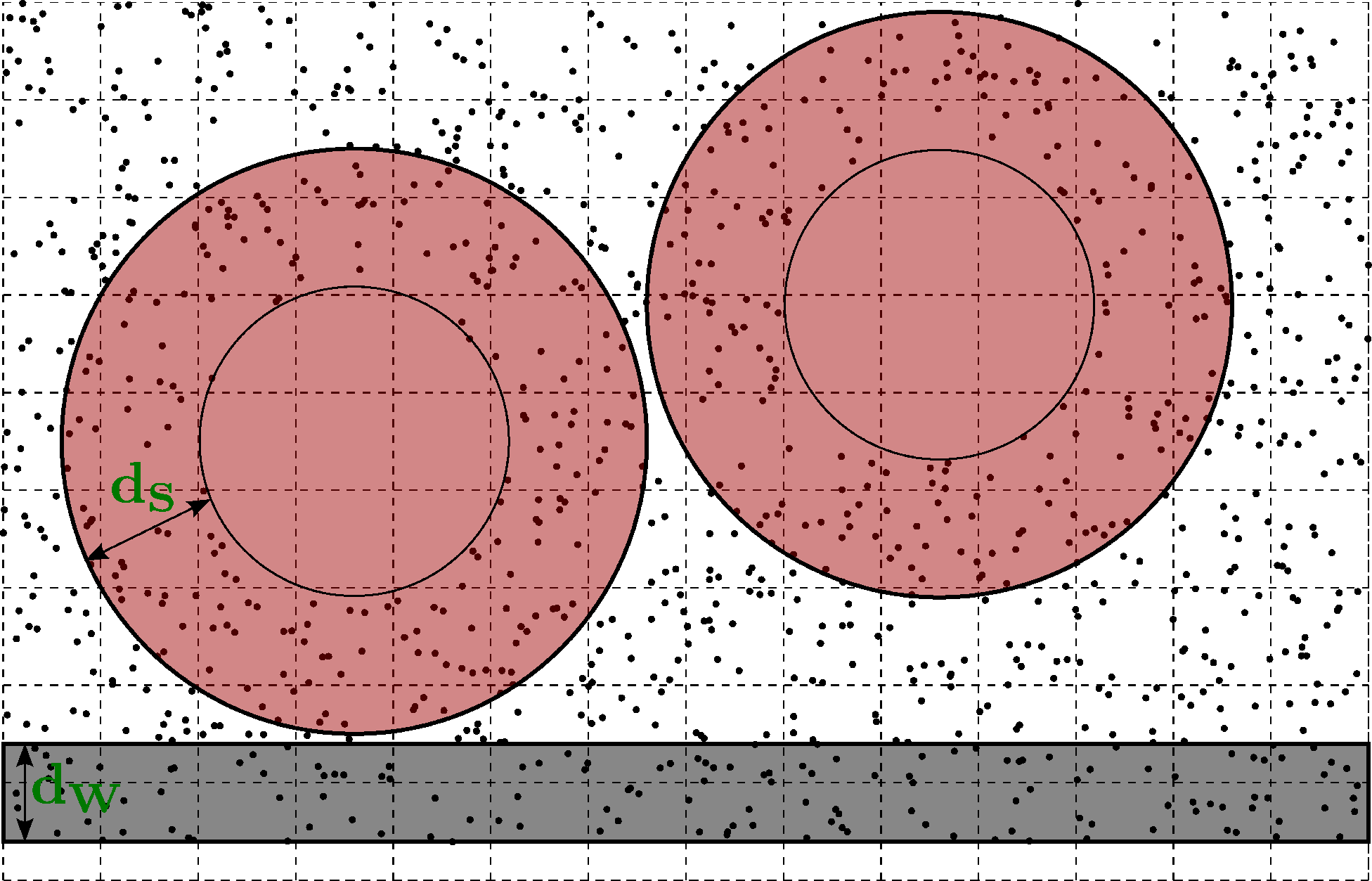}
\end{center}
\caption{2D sketch of  collision step.
Squirmers are shown in red and a wall in grey.
Small dots in the white region indicate the  fluid particles
and dots in the squirmers and in the wall are \textit{virtual} particles,
which are located in a thin layer of thickness $d_W=a_0$ in the wall
and in a thin shell of thickness $d_S = \sqrt{3}a_0$ in the squirmers.
The dashed lines show the collision cell grid.
 }
\label{Fig:vir2}
\end{figure}

In the collision step fluid particles interact with each other but also interact with squirmers and 
bounding
walls 
via
\textit{virtual particles} \cite{Lamura01}.
The collision step starts with shifting
the cell grid
by a random vector $\mathbf{s}$ as discussed in \refSec{MPCD}.
Then, collision cells partly overlapping with squirmers and walls are filled with virtual particles, which
are placed in the walls and in the squirmers,
see \refFig{vir2}.
This increases the accuracy of the 
hydrodynamic flow fields
significantly 
since otherwise these cells
would have an average fluid particle number 
below the mean number $\gamma$
and hence  locally a smaller viscosity \cite{Gompper08}.

The virtual particles are placed  randomly distributed  at density $\rho_f$ in a layer of thickness $d_W=a_0$ inside 
a bounding wall
and in layers of thickness $d_S=\sqrt{3}a_0$ inside the squirmers, see \refFig{vir2}.
It is always guaranteed that
grid cells
overlapping with walls and squirmers
are always completely filled.
The positions and velocities
of the  virtual particles are denoted 
by  $\bar{\mathbf{r}}_{i}$  and $\bar{\mathbf{v}}_{i}$, 
respectively,
with $i=1,\dots, N^v$ and $N^v$ is the total number of virtual particles. 
Their velocities are drawn 
 from a normal distribution with 
 the usual
standard deviation, $\sigma_0=\sqrt{k_BT_0/m_0}$, and zero mean.
In addition, virtual particle $i$ 
located in squirmer $j$ also assumes the local 
slip
velocity of the point $\bar{\mathbf{r}}_{ij}^\ast$  on the squirmer surface, which is closest to $\bar{\mathbf{r}}_i$,
plus the velocity of this point due to the translation and rotation of the squirmer.
So, one sets
\cite{Goetze10}
\begin{equation}
\bar{\mathbf{v}}_{i} = \bar{\mathbf{v}}_{i}^\text{r} + \mathbf{v}_s(\eh_j,\bar{\mathbf{r}}_{ij}^\ast)
  + \boldsymbol{\Omega}_j \times (\bar{\mathbf{r}}_{ij}^\ast-\mathbf{R}_j) + \mathbf{V}_j,
\end{equation}
where $\bar{\mathbf{v}}_{i}^\text{r}$ are the random velocities.
Then, 
as usual,
fluid and virtual particles are sorted into the cells.
We
denote the number of fluid and virtual particles in cell $\xi$ by $N_\xi^f$ and $N_\xi^v$, respectively.
The total number of particles in the cell is $N_\xi=N_\xi^f+N_\xi^v$.
In order to perform 
now
the collision step, the mean velocity in a cell is computed,
\begin{equation}
 \mathbf{u}_\xi = \frac{1}{N_\xi}\left(\sum_{i=1}^{N_\xi^f}\mathbf{v}_i + \sum_{i=1}^{N_\xi^v}\bar{\mathbf{v}}_i\right) \, ,
\end{equation}
and the new velocities of
the particles in a cell 
are set by
the mean velocity
$ \mathbf{u}_\xi$
plus 
a term given by the collision operator, see \refEq{coll000}.

When the Anderson thermostat is used,
random 
velocities have to be computed, which are 
denoted by $\mathbf{v}_i^{\text{r}}$ and $\bar{\mathbf{v}}_i^{\text{r}}$, 
and which are
again drawn from a Gaussian distribution with
width $\sigma_0=\sqrt{k_BT_0/m_0}$.
Then,
in order to conserve linear momentum the change of total velocity due to the added random velocities, 
$\boldsymbol{\mathcal{V}}_\xi$, has to be subtracted.
If conservation of angular momentum is required, by using the collision operator MPC-AT+a, a further term needs to be added.
To implement the collision operator, first
the center of mass has to be calculated, 
\begin{equation}
  \mathbf{r}_\xi^s = \frac{1}{N_\xi}\left(\sum_{i=1}^{N_\xi^f}\mathbf{r}_i + \sum_{i=1}^{N_\xi^v}\bar{\mathbf{r}}_i\right),
\end{equation}
and the relative positions $\mathbf{r}_i^s = \mathbf{r}_i - \mathbf{r}_\xi^s$
and $\bar{\mathbf{r}}_i^s = \bar{\mathbf{r}}_i - \mathbf{r}_\xi^s$.
They are needed to calculate
the inverse of the moment of inertia tensor $\mathbf{I}^{-1}_\xi$ in each cell,
where
\begin{equation}
  \mathbf{I}_\xi = m_0 \left( \sum_{i=1}^{N_\xi^f}(|\mathbf{r}_i^s|^2\,\tEins - \mathbf{r}_i^s\otimes\mathbf{r}_i^s)
         + \sum_{i=1}^{N_\xi^v}(|\bar{\mathbf{r}}_i^s|^2\,\tEins - \bar{\mathbf{r}}_i^s\otimes\bar{\mathbf{r}}_i^s) \right).
\end{equation}
The random velocities added in the collision step change angular momentum by
\begin{equation}
  \Delta \boldsymbol{\mathcal{L}}_\xi = m_0\left( \sum_{i=1}^{N_\xi^f}[\mathbf{r}_i^s\times(\mathbf{v}_i-\mathbf{v}_i^\text{r})]   
   +  \sum_{i=1}^{N_\xi^v}[\bar{\mathbf{r}}_i^s\times(\bar{\mathbf{v}}_i-\bar{\mathbf{v}}_i^\text{r})]  \right).
\end{equation}
To compensate for $ \Delta \boldsymbol{\mathcal{L}}_\xi$ and
conserve angular momentum in each cell, 
the following angular velocity 
is computed \cite{Noguchi07},
\begin{equation}
 \boldsymbol{\omega}_\xi^{\text{AMC}} = \mathbf{I}_\xi^{-1}  \Delta \boldsymbol{\mathcal{L}}_\xi,   
\label{Eq:AT+aOm}
\end{equation}
and used
to rotate 
the
particle velocities in a cell.
Thus, by adding the
extra terms $\boldsymbol{\omega}_\xi^{\text{AMC}} \times \mathbf{r}_i^s$
or $ \boldsymbol{\omega}_\xi^{\text{AMC}} \times \bar{\mathbf{r}}_i^s$
to the new particle velocities, angular momentum is conserved without changing linear momentum.
To summarize, in the collision step the particle
velocities are updated according to \cite{Noguchi07}
\begin{equation}
 \begin{split}
  \mathbf{v}_i' &=  \mathbf{u}_\xi +  \mathbf{v}_i^{\text{r}} - \boldsymbol{\mathcal{V}}_\xi + \boldsymbol{\omega}_\xi^{\text{AMC}} \times \mathbf{r}_i^s, \\
  \bar{\mathbf{v}}_i' &= \mathbf{u}_\xi +  \bar{\mathbf{v}}_i^{\text{r}} - \boldsymbol{\mathcal{V}}_\xi  + \boldsymbol{\omega}_\xi^{\text{AMC}} \times \bar{\mathbf{r}}_i^s \, . \\
 \end{split}
\label{Eq:vAMC+a}
\end{equation}

Finally, momentum \textit{and} angular momentum is transferred to the squirmers,
no matter 
which collision operator 
 has been used.
In each collision cell 
momentum (and angular momentum) is conserved locally
but exchanged between fluid particles and virtual particles.
The change of  momentum for a virtual particle is then  
 \begin{equation}
  \Delta \bar{\mathbf{p}}_i= 
    m_0(\bar{\mathbf{v}}_i'-\bar{\mathbf{v}}_i) .  
\end{equation}
The momenta and angular momenta of all
virtual particles $i$ located in squirmer $j$ are then added
up and assigned
to squirmer $j$.
Thus, after a collision step each
squirmer assumes an additional momentum 
$ \Delta \mathbf{P}_j^c$ and angular momentum $ \Delta \mathbf{L}_j^c$,
 \begin{equation}
 \begin{split}
  \Delta \mathbf{P}_j^c &= \sum_{i=1}^{N_j^v}\Delta \bar{\mathbf{p}}_i, \\
  \Delta \mathbf{L}_j^c &= \sum_{i=1}^{N_j^v} ( \bar{\mathbf{r}}_i - \mathbf{R}_j) \times \Delta\bar{\mathbf{p}}_i,  \\
 \end{split}
\label{Eq:dPCol}
\end{equation}
where the sum goes over all virtual particles   located in squirmer $j$
with total number $N_j^v$.
So, 
after each collision step
the squirmer velocities and angular velocities are updated according to
 \begin{equation}
 \begin{split}
   \mathbf{V}_j' &= \mathbf{V}_j +\frac{1}{M_S} \mathbf{P}_j^c, \\
   \boldsymbol{\Omega}_j' &= \boldsymbol{\Omega}_j +\frac{1}{I_S} \mathbf{L}_j^c \, ,
 \end{split}
\end{equation}
where $I_S$ is the moment of inertia of the squirmer.

\section{Dynamics of a squirmer in a narrow slit geometry}
\label{Sec:slit}
Finally, we apply the discussed simulation procedure to the case of a single squirmer moving in a narrow-slit 
or Hele-Shaw
geometry.
We use a squirmer of radius $R=3a_0$,
swimming speed determined by
$B_1=0.1a_0/\tau_0$, 
and choose different values of $\beta=B_2 / B_1$.
We place 
planar
walls at positions $x=\pm 4a_0$ such that there is only a small gap between squirmer 
surface
and walls.
Then, we perform several simulations at various $\beta$ and initial conditions.

\begin{figure}[bth]
\begin{center}
\includegraphics[width=\columnwidth]{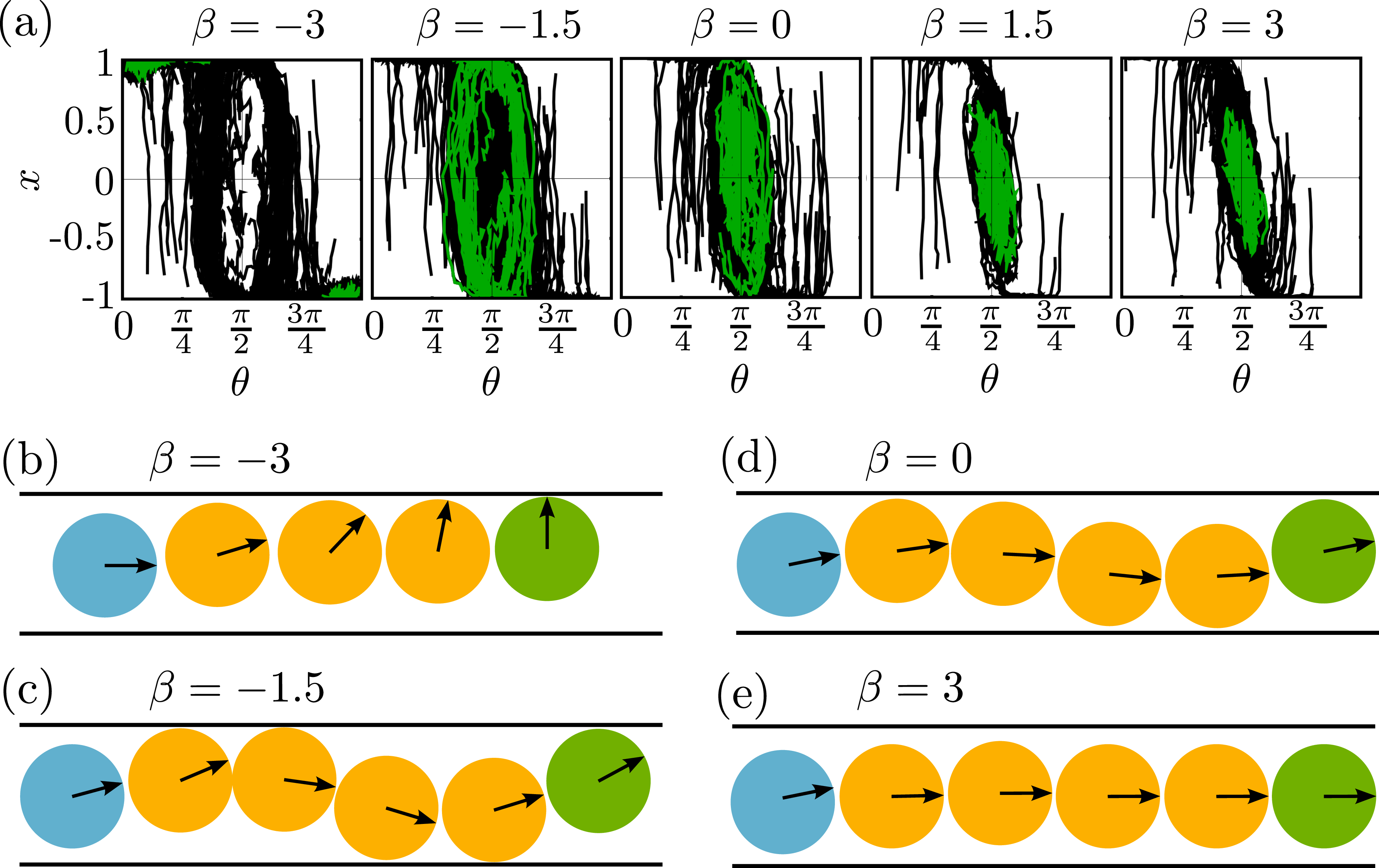}
\end{center}
\caption{
(a) Typical 
phase portraits
from MPCD simulations for a squirmer moving in a narrow-slit geometry.
Walls are located at $x= \pm 4a_0$ and the radius of the squirmer is $R=3a_0$.
Thus,  minimum and maximum positions of its center
are 
at
$x= \pm a_0$. 
The angle $\theta$ 
indicates
the orientation of the squirmer 
relative to the wall normal. So,
$\theta=0$ ($\theta=\pi/2$) is the orientation perpendicular (parallel) to the walls. 
The long-time dynamics is 
shown
in green.
(b-e): Sketch of typical squirmer trajectories for different $\beta$. The initial positions are shown in blue, the final positions in green, and intermediate positions in orange. The orientation vectors are sketched by black arrows.
 }
\label{Fig:slit}
\end{figure}

All the recorded trajectories
are shown in \refFig{slit}(a)
where the position $x$ is plotted versus the angle $\theta$, which is the orientation of the squirmer relative to the upper wall normal.
For 
strongly
negative $\beta$ values (strong pushers) the squirmer tends to orient perpendicular to the walls (shown for $\beta=-3$ in \refFig{slit}(a)).
This is also sketched in \refFig{slit}(b).
Increasing $\beta$ above a certain value induces oscillations of the swimmer between the walls: The squirmer hits a surface, spends some time there, reorients away from the surface, hits the second surface and so on (shown for $\beta=-1.5$ in \refFig{slit}(a) and sketched in \refFig{slit}(c)).
A
neutral squirmer 
still performs oscillations but mainly avoids touching the walls
by swimming with an orientation performing small oscillations about $\theta=\pi/2$ along the bounding walls, see also \refFig{slit}(d).
In contrast, pullers ($\beta >0$) swim stable between the two walls, as sketched in \refFig{slit}(e). Interestingly, the 
$\beta$-dependent
dynamics can be captured by considering the lubrication approximation of a squirmer in front of a wall \cite{Ishikawa06,Schaar2015}, which predicts the correct dynamics for all considered $\beta$ values \cite{Zottl2014}.

\begin{figure}[bth]
\begin{center}
\includegraphics[width=0.8\columnwidth]{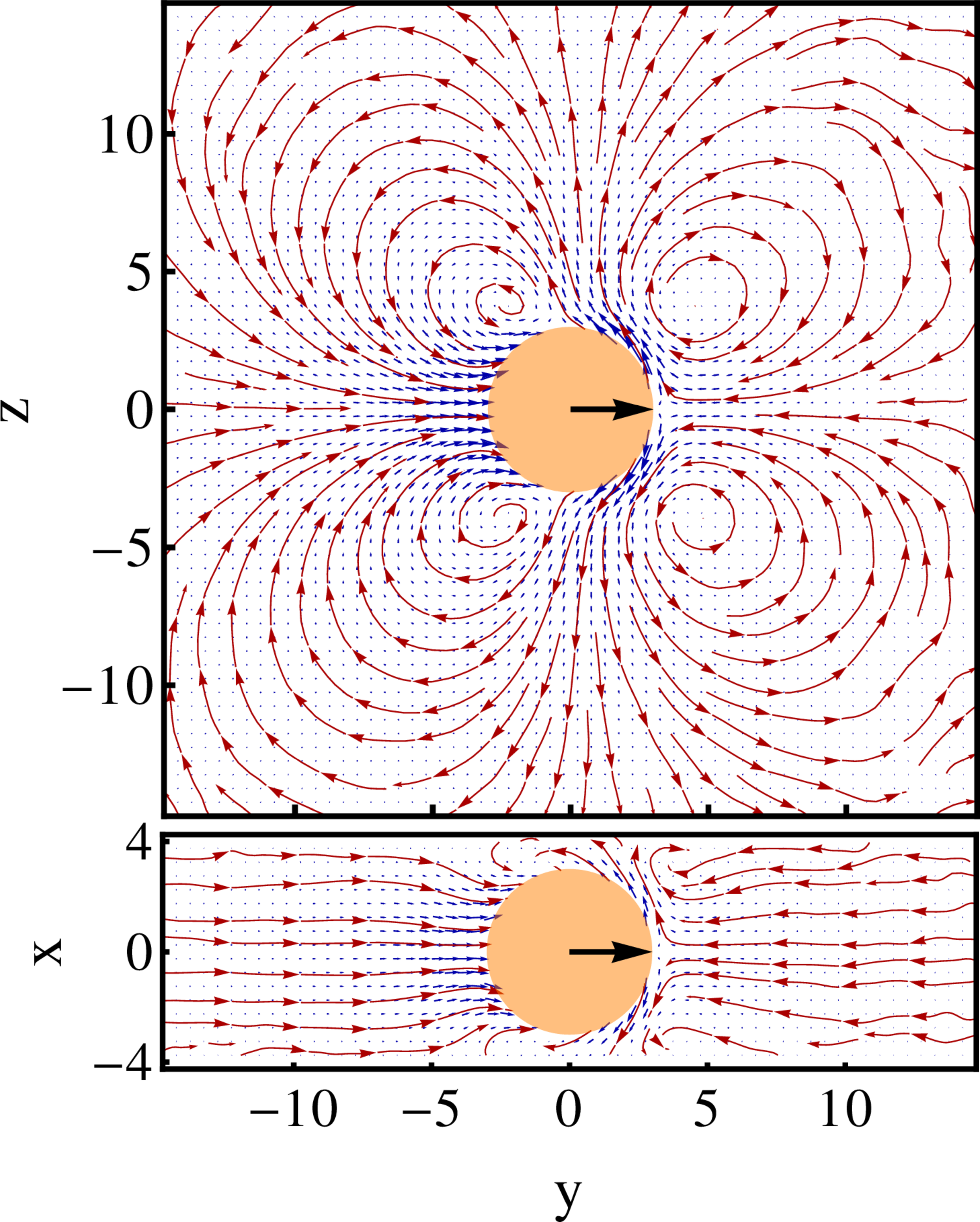}
\end{center}
\caption{
Flow 
field
for a puller swimming in a narrow-slit 
geometry.
The velocity field is indicated in blue and streamlines in red.
Top: Top view of the flow field.
Bottom: side view.
 }
\label{Fig:flow}
\end{figure}

We also 
measured
the swimming velocity and
the flow field around a squirmer with radius $R=3a_0$ and surface velocity modes $B_1=0.1a_0/\tau_0$ and $\beta=3$,
which swims stable between the plates,
by averaging over many realisations and over time.
The swimming velocity results in $v_0^{\text{MPCD}}=0.0657a_0/ \tau_0$, which is very close to the analytic expression for the swimming speed in bulk $v_0=0.0667$ [Eq.~(\ref{Eq:v0})].
To obtain the flow fields,
the simulation box around the swimmer is divided into small cubic boxes, 
for which we determine the flow velocity vector. The boxes
do not have to coincide with collision cell boxes and may even have higher resolution.
\refFigure{flow}  shows the flow field (blue arrows) and stream lines (red) for a slice of the field parallel to the walls (top) and perpendicular to the walls (bottom) in the lab fame.
Due to the presence of the walls the force dipole fields develop closed loops 
\cite{Cisneros07,Hernandez-Ortiz09}.
We use
a grid size with edge length $a_0/2$ to monitor the
velocity field,
meaning that MPCD 
is able
to resolve flow fields below the 
the cell size $a_0$ of the
collision grid.

\begin{figure}[bth]
\begin{center}
\includegraphics[width=\columnwidth]{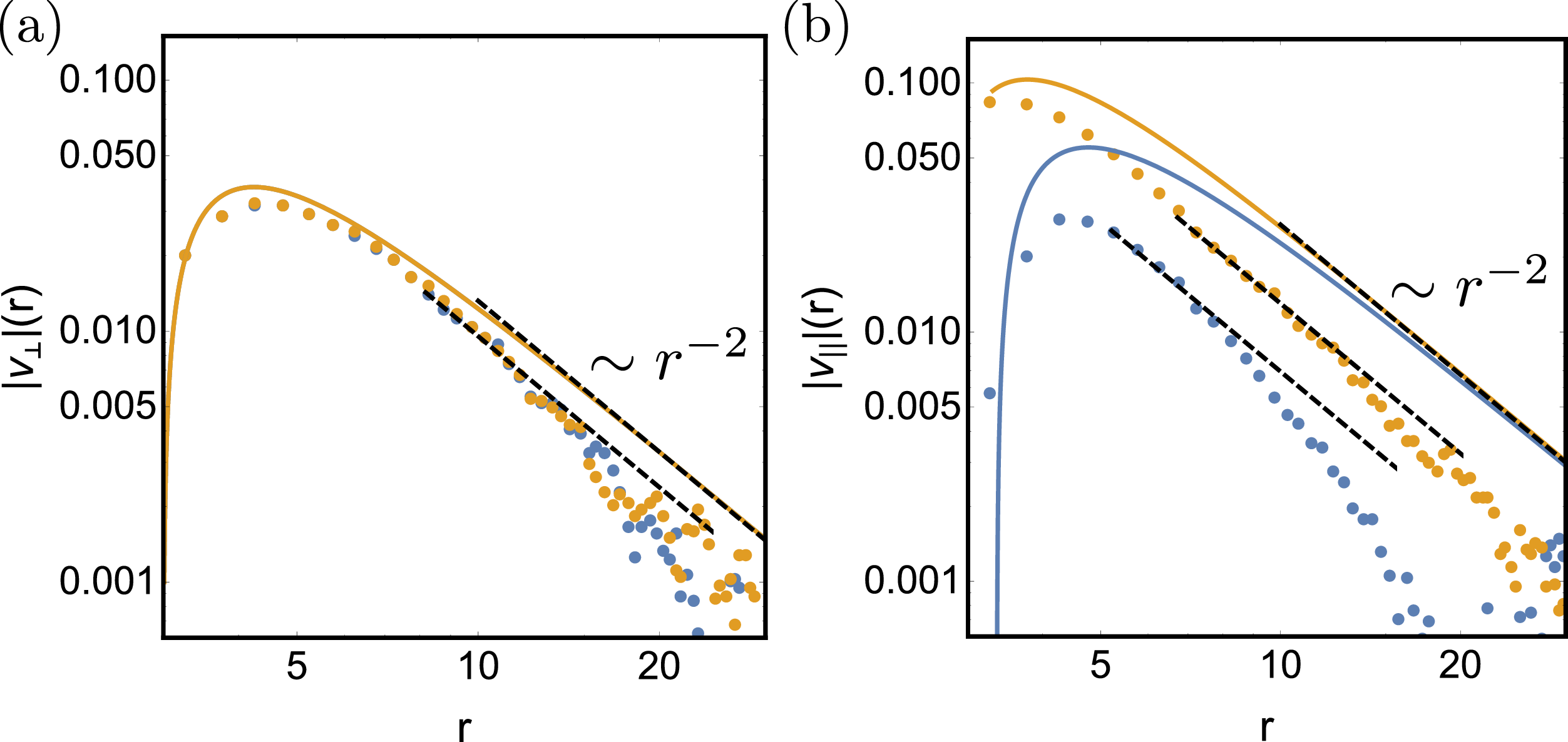}
\end{center}
\caption{
(a) Decay of the flow field perpendicular ($v_\perp$) to the swimmer direction.
The flow field in the $-z$ direction is shown in blue, and in the  $+z$ direction in orange.
(b) Decay of the flow field parallel ($v_{||}$) to the swimmer direction.
The flow field in the $+y$ direction (front of the swimmer) is shown in blue, and in the  $-y$ direction (back of the swimmer) in orange.
Solid lines are curves obtained from the analytic model of a squirmer in bulk [\refEq{VFieldSq1}].
}
\label{Fig:flowdecay}
\end{figure}

In \refFig{flowdecay} we show the decay of the flow field  perpendicular ($v_\perp$) and  parallel ($v_{||}$) to the swimming direction,
which we compare to the decay in bulk using the analytic expressions [\refEq{VFieldSq1}].
Due to the leading force dipole field, they decay with $\sim r^{-2}$.
For intermediate distances away from the squirmer the flow fields obtained from our simulations in strong confinement show a similar decay.
Indeed it has been shown analytically that for very strong confinement flow fields of microswimmers decay in general with  $\sim r^{-2}$ \cite{Brotto2013,Blaschke2016}.
However, in our simulations the decay of the flow fields far away from the squirmer is faster, which probably stems from the periodic boundary conditions \cite{Hu2015,Zottl2017}.
Also, the fact that the strength of the flow field in front of the swimmer is weaker than at the back may  be a signature of the periodic boundary conditions or due to the fact that MPCD does not model a perfect incompressible Stokes flow, but finite Reynolds and Schmidt numbers which account for finite fluid inertia and compressibility, respectively \cite{Padding06,Gompper08}.

\section{Conclusion}
\label{Sec:conc}
Here we have reviewed and discussed how simple spherical model swimmers -- squirmers -- can be modeled and simulated using the method of multiparticle collision dynamics. We have shown how the dynamics of the fluid and the squirmers is coupled both in the streaming step and in the collision step, where a correct momentum and angular momentum transfer between particles ensures  correct hydrodynamics \cite{Downton09} and thermal fluctuations \cite{Goetze10}.
In order to enhance the accuracy of the flow fields around the 
squirmer,
the use of virtual particles inside squirmers, which contribute to the collision step, is necessary.
Also bounding surfaces and pressure-driven flows can be simulated straightforward.
Since the method models the fluid explicitely via effective fluid particles, thermal noise is included naturally in the simulation.
In order to obtain smooth flow fields averaging over several realisations and over time is necessary.



\end{document}